# Asymptotic expression of the virial coefficients for hard sphere systems

by Richard BONNEVILLE

Centre National d'Etudes Spatiales (CNES), 2 place Maurice Quentin, 75001 Paris, France

phone: +33 1 44 76 76 38, fax: +33 1 44 76 78 59, mailto:richard.bonneville@cnes.fr

**Abstract**

We evidence via a computation in the reciprocal space the asymptotic behavior of the high order virial coefficients for a hard sphere system. These coefficients, if their order is high enough, are those of a geometric series. We thus are able to give an explicit expression of the equation of state of the hard sphere system at high density when the fluid phase is no longer the stable one; in the disordered phase this equation of state exhibits a simple pole at the random close packing density. We can then estimate the packing densities of the freezing point of the disordered phase and also of the melting point of the fcc ordered phase. The results are compared with those of the numerical simulations.

## 1. Introduction

The pressure $P$ of a hard sphere fluid can be expressed as an expansion in powers of the packing density $\xi = Nv/V$ as:

$$\frac{Pv}{k_B T} = \xi + B_2\xi^2 + B_3\xi^3 + B_4\xi^4 + ... \quad (1)$$

$k_B$ is the Boltzmann constant, T is the absolute temperature, N the number of particles (which are assumed to be identical), V the system volume, v the molecular volume; N and V are arbitrary large but finite, whereas $\xi$ is a finite parameter.

The virial coefficients $B_p$, either derived from theoretical equations of states, e.g. the celebrated Carnahan and Starling equation [1], or numerically calculated, see Clisby and Mc Coy for the lowest orders (up to 10) [2] or more recently Schultz and Kofke (up to 11) [3], seem to be the terms of a monotonously increasing, absolutely converging series. The radius of convergence derived from the Carnahan and Starling equation of state is 1, however it is well known that the maximum packing density of a system of randomly packed hard spheres is $\xi_0 \cong 0.6366$ [4,5], a value quite close to $2/\pi$. Actually, it has been rigorously demonstrated by Dixmier [6] that this limit is exactly $2/\pi$ when the number of particles tends to infinity and it is worth pointing out that this value of $2/\pi$ is imposed by geometry only and not by thermodynamics. Nevertheless that upper limit of the random close packing density can actually be exceeded by exerting some shearing which creates locally ordered domains [7], and the packing density is then ultimately capped by the maximum fcc packing density $\xi_{fcc} = \pi\sqrt{2}/6 \cong 0.7405$.

That apparent disagreement can be solved if the behavior of the virial coefficients changes for high orders; actually recent extrapolations [8] suggest that the reduced virial coefficients defined by $\overline{B}_p = B_p \xi_0^p$ undergo a peak value for p=14, and could even get negative for p>16, the resulting virial expansion converging for all densities up to the random close packing density where it diverges [9]. We will check that this is indeed the case and we will investigate in more details the behavior of the hard sphere system at high density.

In a recent paper [10], we had presented a new analytical approach to derive the virial coefficients for simple fluids with a focus on the hard sphere fluid case; we had shown how working in the reciprocal space may facilitate the computation of those coefficients which are then expressed as multiple integral expressions in the reciprocal space.

In this paper we will give an explicit asymptotic expression of the virial coefficients and of the equation of state of the hard sphere system in the disordered phase at high density when the fluid phase is no longer the stable phase. We will then estimate the densities of the freezing point of the disordered phase and of the melting point of the fcc ordered phase, and compare the results with those of the numerical simulations.

## 2. The virial coefficient computation in the reciprocal space

In this section we summarize the process detailed in ref.[9]. The starting point is the expression of the semi classical partition function:

$$\mathbb{Z} \cong \frac{(2\pi M k_B T)^{3N/2}}{\hbar^{3N}} \frac{V^N}{N!} Z, \tag{2a}$$

where

$$Z = V^{-N} \int_{(V)} d^3 \mathbf{R}_1 \int_{(V)} d^3 \mathbf{R}_2 ... \int_{(V)} d^3 \mathbf{R}_N \exp\left(\frac{-1}{k_B T} \sum_{p>q} U(\mathbf{R}_p - \mathbf{R}_q)\right). \tag{2b}$$

$U(\mathbf{R}_p - \mathbf{R}_q)$ is the intermolecular potential, assuming only pair interactions depending upon the relative position of the molecules.

We first perform a Fourier transformation of the expression equ.(2b):

$$\exp - U(\mathbf{R}_p - \mathbf{R}_q)/k_B T = \int \frac{d^3 \mathbf{k}_{pq}}{(2\pi)^3} \left[(2\pi)^3 \delta(\mathbf{k}_{pq}) + \Phi(\mathbf{k}_{pq})\right] \exp\left(i\mathbf{k}_{pq}.(\mathbf{R}_p - \mathbf{R}_q)\right) \tag{3}$$

From equ.(2b) and equ.(3) Z can then be written as:

$$Z = \prod_{p>q} \int \frac{d^3\mathbf{k}_{pq}}{(2\pi)^3} \left[ (2\pi)^3 \delta(\mathbf{k}_{pq}) + \Phi(\mathbf{k}_{pq}) \right] \Delta \tag{4}$$

We have put

$$\Delta = \prod_m \int_{(V)} \frac{d^3\mathbf{R}_m}{V} \exp\left( i\mathbf{R}_m \cdot \sum_n \mathbf{k}_{mn} \right) \text{ with } \mathbf{k}_{mn} = -\mathbf{k}_{nm} \tag{5}$$

In the limit $V \to \infty$,

$$\Delta = \prod_m \frac{(2\pi)^3}{V} \delta\left( \sum_n \mathbf{k}_{mn} \right) \tag{6}$$

$Z$ can then be expanded as a series whose term of order n is a multiple integral involving $\frac{\Phi}{V}$ n times. A systematic procedure involving combinatorial analysis and partial re-summations finally allows expressing Z as

$$Z = \exp N \left( \frac{NI_2}{2!} + \frac{N^2 I_3}{3!} + \frac{N^3 I_4}{4!} + \frac{N^4 I_5}{5!} + \ldots \right) \tag{7}$$

Each $I_p$ is the sum of a set of integrals in the reciprocal space (see in Appendix A1 the examples of $I_2, I_3, I_4$) such as

$$I_p^{[m_n]} = \int \ldots \int \frac{\Phi_1}{V} \frac{\Phi_2}{V} \ldots \frac{\Phi_{p+m-1}}{V} \frac{V d^3\mathbf{k}_1}{(2\pi)^3} \frac{V d^3\mathbf{k}_2}{(2\pi)^3} \ldots \frac{V d^3\mathbf{k}_m}{(2\pi)^3} \tag{8}$$

The index m designates the number of independent integration variables (N.B. the notation is slightly different from the notation used in ref.[10]). The additional sub-index $n$ allows to discriminate among the possible several distinct integrals corresponding to the same values of p and m.

The integrals $I_p^{[m_n]}$ are practically obtained by considering a multiplet of p particles and all the possible anti-symmetric p×p matrices { $\mathbf{k}_{ij}$ }, i and j ranging from 1 to p, such that the sum of all the elements on a same line (or column) is null. The $\mathbf{k}_{ij}$ may or may not be identically null. An integral $I_p^{[m_n]}$ is associated to each distinct type of matrix, but the same integral corresponds to distinct matrices differing only by index permutations, which is accounted for by a multiplicity factor $g_p^{[m_n]}$. It finally results in the effective contribution of order p

$$I_p = \sum_{m,n} g_p^{[m_n]} I_p^{[m_n]} \qquad (9)$$

From $P = k_B T \partial \text{Log} \mathbb{Z} / \partial V$ we derive the equation of state as

$$\frac{P}{k_B T} = \frac{N}{V} - \frac{N^2}{2!V} I_2 - \frac{2N^3}{3!V} I_3 - \frac{3N^4}{4!V} I_4 - \frac{4N^5}{5!V} I_5 + ... \qquad (10)$$

We have here above assumed no specific form for the intermolecular potential; we will now focus on the hard sphere case. For hard spheres of diameter d and volume $v = \pi d^3 / 6$

$$\Phi(\mathbf{k}) = -\frac{4\pi}{k^3} \left( \sin(kd) - kd \cos kd \right) \qquad (11a)$$

Let us introduce the dimensionless variables $\mathbf{q}_i = \mathbf{k}_i d$ and the reduced functions

$$\varphi(q_i) = -\frac{24}{q_i^3} \left( \sin(q_i) - q_i \cos(q_i) \right) \qquad (11b)$$

with $q = |\mathbf{q}|$. $I_p^{[m_n]}$ can then be written as $(v/V)^{p-1}$ times a dimensionless, purely numerical, factor:

$$I_p^{[m_n]} = \left( \frac{v}{V} \right)^{p-1} \left( \frac{\pi}{6} \right)^m \iiint \varphi_1 \varphi_2 ... \varphi_{-m+p(p-1)/2} \frac{d^3\mathbf{q}_1}{(2\pi)^3} \frac{d^3\mathbf{q}_2}{(2\pi)^3} ... \frac{d^3\mathbf{q}_m}{(2\pi)^3} \qquad (12)$$

The equation of state equ.(10) is written as an expansion of the packing density $\xi = Nv/V$ in the form of equ.(1), the virial coefficients $B_p$ being connected to the integrals $I_p$ by the relation

$$B_p = -\frac{(p-1)}{p!} \left( \frac{V}{v} \right)^{p-1} I_p \qquad (13)$$

The Carnahan & Starling equation of state can be derived from the assumption of a recursion relation between the $I_p$ s, namely

$$\frac{I_{p+1}}{I_p} = \frac{(p+3)(p+1)}{p+2}\left(\frac{v}{V}\right) \qquad \text{i.e.} \qquad I_p = -(p+2)p!\left(\frac{v}{V}\right)^{p-1} \qquad (14a)$$

which leads to

$$B_p = (p+2)(p-1) \qquad (14b)$$

but there is no evidence that such a relation exists and actually it is not the case.

Among the $I_p^{[m_n]}$, two extreme cases are remarkable:

(i) There are only two non-zero elements $+q$ and $-q$ on every line or column of the matrix $\{\mathbf{k}_{ij}\}$; thus there is only one integration variable, i.e. m=1. For $p \geq 4$ there are

$$g_p^{[1]} = \frac{(p-1)!}{2} \qquad (15)$$

distinct matrices of that form which are associated to the integral

$$I_p^{[1]} = \int \left(\frac{\Phi(q)}{V}\right)^p \frac{V d^3\mathbf{q}}{(2\pi)^3} \qquad (16a)$$

or

$$I_p^{[1]} = \left(\frac{v}{V}\right)^{p-1}\left(\frac{\pi}{6}\right)\int \varphi(q)^p \frac{d^3\mathbf{q}}{(2\pi)^3} \qquad (16b)$$

When $p \to \infty$, it appears <u>numerically</u> that

$$\lim_{p\to\infty} \frac{I_{p+1}^{[1]}}{I_p^{[1]}} \cong -\frac{5\pi}{2}\frac{v}{V} \cong -7.854\frac{v}{V}. \qquad (17)$$

Extending this recursion formula to low values of p gives

$$I_p^{[1]} \cong (-1)^p \frac{\left(\frac{5\pi}{2}\right)^{p-1}}{p-1}\left(\frac{v}{V}\right)^{p-1} \qquad (18)$$

It can be checked that this expression actually is valid even for low values of p.

(ii) None of the $\mathbf{k}_{ij}$ is identically null; since there are $p(p-1)/2$ variables $\mathbf{k}_{ij}$ linked by $(p-1)$ independent conditions, there are $Q=(p-1)(p-2)/2$ independent variables of integration whereas the function $\Phi$ appears $p(p-1)/2$ times in the integrand. There is only one type of matrix of that form and a single integral $I_p^{[Q]}$ whose multiplicity is thus

$$g_p^{[Q]} = 1 \tag{19}$$

The integral $I_p^{[Q]}$ is expressed as

$$I_p^{[Q]} = \int \ldots \int \frac{\Phi_1(\mathbf{k}_1)}{V} \frac{Vd^3\mathbf{k}_1}{(2\pi)^3} \frac{\Phi_2(\mathbf{k}_2)}{V} \frac{Vd^3\mathbf{k}_2}{(2\pi)^3} \ldots \frac{\Phi_Q(\mathbf{k}_Q)}{V} \frac{Vd^3\mathbf{k}_Q}{(2\pi)^3} \frac{\Phi_{Q+1}}{V} \frac{\Phi_{Q+2}}{V} \ldots \frac{\Phi_{p(p-1)/2}}{V} \tag{20}$$

It will be estimated in the next section.

## 3. Asymptotic value of the integrals $I_p^{[m_n]}$

If we consider for a given value of p the terms $I_p^{[m_n]}$ which contain a large number of functions $\Phi$ in the integrand together with a large number of integration variables, the oscillating character of the $\Phi$ functions generates destructive interferences between those functions unless their arguments are identical. If p is high enough we can thus approximate $I_p^{[Q]}$ as in equ.(28) by

$$I_p^{[Q]} \cong \int \left(\frac{\Phi_1(\mathbf{k}_1)}{V}\right)^p \frac{Vd^3\mathbf{k}_1}{(2\pi)^3} \left[\int \ldots \int \frac{\Phi_2}{V} \ldots \frac{\Phi_Q}{V} \frac{Vd^3\mathbf{k}_2}{(2\pi)^3} \ldots \frac{Vd^3\mathbf{k}_Q}{(2\pi)^3}\right] \tag{21}$$

In the bracketed integral we can check that the number $\frac{(p-1)(p-2)}{2}-1$ of integration variables $\mathbf{k}$ and the number $\frac{p(p-1)}{2}-p$ of functions $\Phi$ are the same i.e. $\frac{p(p-3)}{2}=Q-1$.

Moreover when $p \to \infty$ the number of variables, varying as $p^2/2$, becomes larger and larger,

much larger than the number of relations, varying as p, linking the arguments of the functions so that the integrations tend to be independent from each other; we can write in that limit

$$I_p^{[Q]} \cong \int \left(\frac{\Phi_1(\mathbf{k}_1)}{V}\right)^p \frac{Vd^3\mathbf{k}_1}{(2\pi)^3} \left(\int \frac{\Phi_2}{V} \frac{Vd^3\mathbf{k}_2}{(2\pi)^3}\right)^{\frac{p(p-3)}{2}} \tag{22}$$

Now we have (see Appendix A2)

$$\int \frac{\Phi(k)}{V} \frac{Vd^3k}{(2\pi)^3} = -1 \tag{23}$$

We deduce

$$\begin{aligned}I_p^{[Q]} &\cong (-1)^{\frac{p(p-3)}{2}} \int \left(\frac{\Phi_1(\mathbf{k}_1)}{V}\right)^p \frac{Vd^3\mathbf{k}_1}{(2\pi)^3} \\ &= (-1)^{\frac{p(p-3)}{2}} I_p^{[1]}\end{aligned} \tag{24}$$

As we had previously seen that $I_p^{[1]} \cong (-1)^p \dfrac{\left(\dfrac{5\pi}{2}\right)^{p-1}}{p-1}\left(\dfrac{v}{V}\right)^{p-1}$ we find that for high enough values of p the $I_p^{[Q]}$s are well accounted for by the formula

$$I_p^{[Q]} \cong (-1)^{\frac{p(p-1)}{2}} \frac{\left(\frac{5\pi}{2}\right)^{p-1}}{p} \left(\frac{v}{V}\right)^{p-1} \tag{25}$$

We infer from equ.(19) and equ.(25) that for high enough values of p

$$I_p^{[m]} \cong (-1)^{m+p-1} \frac{\left(\frac{5\pi}{2}\right)^{p-1}}{p} \left(\frac{v}{V}\right)^{p-1} \tag{26}$$

Thus we see that the different integrals associated to a given, high enough value of p tend toward a same absolute value but are alternatively positive and negative according to the number of integration variables.

## 4. Asymptotic value of the multiplicity factors $g_p^{[m_n]}$

Now we have to take into account the multiplicity factors $g_p^{[m_n]}$. Let us consider a polyhedron with p summits inside a space with $(p-1)$ dimensions; the summits represent the particles, the links between two summits represent the $\mathbf{k}_{ij}$s. As in section 2 & 3 above, there are two special situations:

(i) the situation where every summit is connected to all the other summits, i.e. all its elements $i \neq j$ of the $\{\mathbf{k}_{ij}\}$ anti-symmetric matrix are non-zero: there is only one possible configuration and it illustrates the case $g_p^{[Q]} = 1$.

(ii) the situation where every summit is only connected to 2 other summits, so realizing a closed loop that can be gone over without passing twice over the same link; it illustrates the case $g_p^{[1]} = \dfrac{(p-1)!}{2}$.

We can start from the 1$^{st}$ situation and delete one link in the figure, i.e. one of the $\mathbf{k}_{ij}$s, e.g. $\mathbf{k}_{12}$, is null. Alternatively we can start from the 2$^{nd}$ situation and add one link in the figure. We can repeat this process so as to gradually obtain two rapidly increasing sequences which join together at a common peak value for a value $\overline{m}$ of the index $m_n$. The effective counting of the multiplicity of the successive terms is an increasingly cumbersome task but fortunately we do not need it. $\overline{m}$ is of the order of the middle of the interval $[2, Q]$ spanned by $m_n$, i.e.

$$\overline{m} \approx Q/2 \cong p^2/4 \tag{27a}$$

Since $p \gg 1$ we can discard the odd or even character of p. This can be re-written as

$$\overline{m} \approx \frac{1}{2} p \times \frac{p}{2} \tag{27b}$$

It means that for the peak value $g_p^{[\overline{m}]}$ each of the p summits is connected in average with $\approx p/2$ other summits (the factor 1/2 is there not to count every configuration twice). Now it appears from considering the expressions $g_p^{[1]} = \frac{(p-1)!}{2}$ and $g_p^{[Q]} = 1$ that the two sequences are not symmetrical and that the common peak is shifted toward the lower values of the index $m_n$, i.e. $\overline{m} \approx Q/2$ but $\overline{m} < Q/2$. We will thus write

$$\overline{m} \cong \frac{1}{2} p \times \frac{p}{r} = \frac{p^2}{2r} \tag{28}$$

It means that each of the p summits is connected in average with p/r other summits with $r \approx 2$ but $r > 2$.

Moreover as p>>1 the peak $g_p^{[\overline{m}]}$ is so much higher than the other $g_p^{[m_n]}$s that we can neglect the contribution of the later and only take $g_p^{[\overline{m}]}$ into account.

By adding one summit to the polyhedron, we pass from the most likely configuration with p summits to the most likely configuration with p+1 summits by the following recursion:

$$g_{p+1} \cong \frac{1}{2} g_p \frac{(p+1)! \, r^p}{r^{p+1} \, p!} \tag{29a}$$

i.e.

$$g_{p+1} = \frac{1}{2} g_p \times \frac{p+1}{r} \cong \frac{1}{2} g_p \times \frac{p}{r} \tag{29b}$$

It means that each of the p+1 summits is connected in average with p/r other summits (the factor 1/2 has been introduced not to count every configuration twice).

5. **Consequence: asymptotic value of the virial coefficients**

By combining equ.(26) and equ.(29b) with $m = \overline{m}$ for $I_p$ and $m = \overline{m}+1$ for $I_{p+1}$, we obtain for the integrals

$$\frac{I_{p+1}}{I_p} \cong \left(\frac{5\pi}{2}\right)\left(\frac{v}{V}\right)\frac{g_{p+1}}{g_p} \cong p\left(\frac{5\pi}{4r}\right)\left(\frac{v}{V}\right) \tag{30}$$

Then by combining equ.(13) and equ.(30) we obtain for the virial coefficients

$$\lim_{p\to\infty}\frac{B_{p+1}}{B_p} \cong \frac{5\pi}{4r} \tag{31}$$

We find that the high order virial coefficients of the hard sphere system are the terms of an absolutely converging series whose radius of convergence determines an upper packing density limit $\xi_0 < 1$ such that

$$\xi_0 = \lim_{p\to\infty}\frac{B_p}{B_{p+1}} = \frac{4r}{5\pi} \tag{32}$$

We have previously said that $r \approx 2$ but also that $r > 2$; an upper limit is given by reminding that in any case $\xi$ is bounded by the fcc close packing density: $\xi_0$ thus ranges between 0.5093 and 0.7405 (see table below). We know that the effective density limit of the disordered phase is the random close packing density $\xi_0 = 2/\pi$ [6], that value being imposed by geometry only as it has been pointed out in the introduction. We conclude that $r$ must be equal to 5/2.

|  | $\xi_0$ | **r** |
|---|---|---|
| **lower limit** | 0.5093 | 2 |
| **random close packing** | $2/\pi \cong 0.6366$ | 5/2 |
| **upper limit (fcc)** | $\pi\sqrt{2}/6 \cong 0.7405$ | 2.908 |

The transition point between this asymptotic regime and the virial expansion at lower density can be estimated as follows. If we extrapolate the recursion relation equ.(29b) to low values of p we get

$$g_p \cong \frac{p!}{(2r)^{p-1}} g_1 \tag{33}$$

Such an extrapolation must be viewed as an unphysical mathematical trick since the recursion relation equ.(37b) does not hold for low values of p and moreover we know that p must be greater than 3. A priori $g_1$ is of the order of unity and will be precisely determined in the next section. Approximating p! by $\left(\frac{p}{e}\right)^p$ we obtain $g_p \approx 2r\left(\frac{p}{2er}\right)^p g_1$. The transition point can be estimated by $\frac{p}{2er} \approx 1$. With $r = 5/2$ we get $p \approx 13$ to 14, as suggested in ref.[7].

As for the virial coefficients, combining equ.(13), equ.(31) and equ.(33) we find the asymptotic value

$$B_p \cong \left(\frac{5\pi}{4r}\right)^{p-1} g_1 \tag{34a}$$

i.e.

$$B_p \cong \frac{g_1}{\xi_0^{p-1}} \tag{34b}$$

Consequently the virial expansion takes the form of a geometric series and the equation of state has the following asymptotic behavior:

$$\frac{Pv}{k_B T} \cong g_1 \frac{\xi}{1 - \xi/\xi_0} \tag{35}$$

As expected, this equation of state exhibits a simple pole for $\xi = \xi_0$.

## 6. Discussion

A sketch of the phase diagram of the hard sphere system can be found e.g. in ref.[4]. Freezing of the fluid phase can occur when the packing density exceeds a value $\xi_f \approx 0.50$; between $\xi_f$ and $\xi_0$ the fluid phase is metastable. Above $\xi_0$ only the ordered phase is stable, its density between ultimately capped by the fcc packing density $\xi_{fcc}$; melting of the ordered phase occurs when its density goes down below $\xi_m \approx 0.55$.

The above equation of state equ.(35) accounts for the asymptotic behavior of the hard sphere system in the disordered phase at high density i.e. above the freezing point where the fluid phase is no longer stable. In the ordered stable phase, the equation of state can be expressed in a quite similar manner to equ.(35) by replacing the maximum random close packing density $\xi_0 = 2/\pi$ by the maximum fcc packing density $\xi_{fcc} = \pi\sqrt{2}/6$ [11]:

$$\frac{Pv}{k_B T} \cong g_1^{fcc} \frac{\xi}{1-\xi/\xi_{fcc}}. \tag{36}$$

Those asymptotic expressions equ.(35) and equ.(36) are a priori valid only in the vicinity of $\xi_0$ and $\xi_{fcc}$ respectively. Extending those expressions toward lower densities, $\xi_f$ for the disordered phase and $\xi_m$ for the ordered phase, can be seen as a rather crude approximation but in fact we will see it works well enough for estimating the densities of the freezing and melting points. We thus assume that equ.(36) is valid in the ordered phase down to the phase transition, i.e. from $\xi_{fcc}$ to $\xi_m$, and correlatively that equ.(35) is also valid on the disordered phase down to the phase transition, i.e. from $\xi_0$ to $\xi_f$.

In the disordered phase, the Carnahan and Starling equation of state [1]

$$\frac{Pv}{k_B T} \cong \xi \frac{1+\xi+\xi^2-\xi^3}{(1-\xi)^3}. \tag{37}$$

is known to work remarkably below the freezing point at low and medium density when the fluid phase is the stable one, and even still slightly beyond in the metastable region. Now, the numerical simulations indicate that on the disordered branch the passing from the stable region to the metastable one is smooth and shows no accident such as a discontinuity or an angular point in the equation of state. Hence there is continuity of the pressure $P(\xi)$ and of its first derivative $\frac{\partial P}{\partial \xi}$ which is the isothermal compressibility. That explains why the Carnahan & Starling equation of state may work well above the freezing point even if we choose to represent the stable phase by the Carnahan & Starling equation and the metastable phase by the asymptotic equation equ.(35). A discontinuity at the freezing point density would only be at the level of the second derivative of the pressure.

### 6.1 Disordered phase

We will express that the free energy F has the same value in both stable and metastable phases for $\xi = \xi_f$ and that there is continuity of $\frac{\partial F}{\partial \xi}$ (i.e. of the pressure) and of $\frac{\partial^2 F}{\partial \xi^2}$ (i.e. of the isothermal compressibility), respectively taking for the equation of state below and above the freezing point the expressions the Carnahan & Starling equation of state equ.(37) and the asymptotic expression equ.(35).

For the stable phase we have

$$F - Nf_0(T) = Nk_B T \left( Log(\xi) - 1 \right) + Nk_B T \sum_{p=2}^{\infty} (p+2)\xi^{p-1}$$
$$= Nk_B T \left( Log(\xi) - 1 + \frac{\xi(4-3\xi)}{(1-\xi)^2} \right) \quad (38a)$$

and

$$\frac{\partial F}{\partial \xi} = Nk_B T \frac{1 + \xi + \xi^2 - \xi^3}{\xi(1-\xi)^3} \quad (38b)$$

$$\frac{\partial^2 F}{\partial \xi^2} = Nk_B T \frac{-1 + 4\xi + 4\xi^2 - \xi^4}{\xi^2(1-\xi)^4} \quad (38c)$$

In equ.(46a) the term $f_0(T) = k_B T Log\left(\dfrac{\hbar^3}{(2\pi M k_B T)^{3/2} v}\right)$ comes from the impulsion degrees of freedom, which are not relevant here, whereas the term (-1) comes from the indistinguishability of the molecules.

For the metastable phase we obtain

$$F - Nf_0(T) = Nk_B Tg_1 Log(\xi/\xi_c) + Nk_B Tg_1 \sum_{p=1}^{\infty} \frac{(\xi/\xi_0)^p}{p}$$
$$= Nk_B Tg_1 Log(\xi/\xi_c) - Nk_B Tg_1 Log(1 - \xi/\xi_0) \tag{39a}$$

where $\xi_c$ is an integration constant, and

$$\frac{\partial F}{\partial \xi} = Nk_B Tg_1 \frac{1}{\xi(1 - \xi/\xi_0)} \tag{39b}$$

$$\frac{\partial^2 F}{\partial \xi^2} = Nk_B Tg_1 \frac{-1 + 2\xi/\xi_0}{\xi^2 (1 - \xi/\xi_0)^2} \tag{39c}$$

$g_1$, $\xi_f$ and $\xi_c$ can now be calculated and we obtain:

$$\xi_f = 0.4855 \tag{40a}$$

$$g_1 = 2.8005 \tag{40b}$$

and

$$\xi_c = 0.7153 \tag{40c}$$

That value of the freezing point density $\xi_f = 0.4855$ is slightly lower (by 2%) than the Monte Carlo result $\xi_f = 0.494$ obtained by Hoover and Ree [12]; correlatively the value of $g_1$ is slightly above the one $(g_1 = 2.765)$ proposed by Speedy [11] and Robles et al. [13]. The agreement is good if we consider the rudeness of the approximation consisting in extending the asymptotic behavior of the equation of state down to the freezing point.

### 6.2 Ordered phase

The agreed value for $g_1^{fcc}$ is 3, i.e. the dimensionality of the system [14], a theoretical result which is confirmed by numerical simulations [15, 16]; it is then possible to estimate the density $\xi_m$ of the melting point. We just have to express that at equilibrium the pressure $P = \left(\frac{\partial F}{\partial V}\right)_N$ and the chemical potential $\mu = \left(\frac{\partial F}{\partial N}\right)_V$ have the same value in the ordered phase $(\xi = \xi_m)$ and the disordered phase $(\xi = \xi_f)$. From the expression of the free energy equ.(39a) we derive the expression of the chemical potential, which is identical to the free enthalpy per molecule $\frac{F + PV}{N}$; for the disordered phase:

$$\frac{\mu}{k_B T} = \frac{f_0(T)}{k_B T} + g_1 Log\left(\frac{\xi/\xi_c}{1 - \xi/\xi_0}\right) + g_1 \frac{1}{1 - \xi/\xi_0} \tag{41a}$$

and for the ordered phase:

$$\frac{\mu}{k_B T} = \frac{f_0(T)}{k_B T} + g_1^{fcc} Log\left(\frac{\xi/\xi_c^{fcc}}{1 - \xi/\xi_c^{fcc}}\right) + g_1^{fcc} \frac{1}{1 - \xi/\xi_c^{fcc}} \tag{41b}$$

With $\xi_f = 0.4855$, $g_1 = 2.8005$, $\xi_c = 0.7153$ and $g_1^{fcc} = 3$ we obtain

$$\xi_m = 0.5336 \tag{42a}$$

and

$$\xi_c^{fcc} = 0.5025 \tag{42b}$$

Once again that value of the melting point density $\xi_m = 0.5336$ is slightly lower (by 2%) than the Monte Carlo result $\xi_m = 0.545$ obtained by Hoover and Ree [12]. As in the case of the freezing point, the agreement is good if we consider the rudeness of the approximation consisting in extending the asymptotic behavior of the equations of state down to the phase transition region.

## 7. Conclusion

Thanks to a computation in the reciprocal space we have shown that the virial coefficients, if their order is high enough, are those of a geometric series. We have thus been able to give an explicit expression of the equation of state of the hard sphere system in the disordered phase at high density when the fluid phase is no longer the stable phase; this equation of state exhibits a simple pole at the random close packing density. We have also been able to estimate the packing densities of the freezing point of the disordered phase and of the melting point of the fcc ordered phase with a good enough agreement with the results of numerical simulations.

Appendix A1

For example we have

$$I_2 = \frac{\Phi(0)}{V}$$

$$I_3 = \int \frac{V d^3\mathbf{k}}{(2\pi)^3} \left(\frac{\Phi(|\mathbf{k}|)}{V}\right)^3$$

$I_4$ is actually the sum of 3 integral terms

$$I_4 = I_4^{[3]} + 6 I_4^{[2]} + 3 I_4^{[1]},$$

with

$$I_4^{[3]} = V^{-3}(2\pi)^{-9} \int\int\int \Phi(|\mathbf{k}_1|)\Phi(|\mathbf{k}_2|)\Phi(|\mathbf{k}_3|)\Phi(|\mathbf{k}_1-\mathbf{k}_2|)\Phi(|\mathbf{k}_2-\mathbf{k}_3|)\Phi(|\mathbf{k}_3-\mathbf{k}_1|) d^3\mathbf{k}_1 d^3\mathbf{k}_2 d^3\mathbf{k}_3$$

$$I_4^{[2]} = V^{-3}(2\pi)^{-6} \int\int \Phi(|\mathbf{k}_1|)^2 \Phi(|\mathbf{k}_2|)^2 \Phi(|\mathbf{k}_1-\mathbf{k}_2|) d^3\mathbf{k}_1 d^3\mathbf{k}_2$$

$$I_4^{[1]} = V^{-3}(2\pi)^{-3} \int \Phi(|\mathbf{k}_1|)^4 d^3\mathbf{k}_1$$

$I_4^{[1]}$ is associated to the 3 matrices

$$\begin{pmatrix} 0 & 0 & q & -q \\ 0 & 0 & -q & q \\ -q & q & 0 & 0 \\ q & -q & 0 & 0 \end{pmatrix}, \begin{pmatrix} 0 & q & 0 & -q \\ -q & 0 & q & 0 \\ 0 & -q & 0 & q \\ q & 0 & -q & 0 \end{pmatrix}, \begin{pmatrix} 0 & q & -q & 0 \\ -q & 0 & 0 & q \\ q & 0 & 0 & -q \\ 0 & -q & q & 0 \end{pmatrix}$$

$I_4^{[3]}$ is associated to the single matrix:

$$\begin{pmatrix} 0 & \mathbf{k}_{12} & \mathbf{k}_{13} & \mathbf{k}_{14} \\ \mathbf{k}_{21} & 0 & \mathbf{k}_{23} & \mathbf{k}_{24} \\ \mathbf{k}_{31} & \mathbf{k}_{32} & 0 & \mathbf{k}_{34} \\ \mathbf{k}_{41} & \mathbf{k}_{42} & \mathbf{k}_{43} & 0 \end{pmatrix} \Rightarrow \begin{pmatrix} 0 & q_1 & -q_2 & -q_1+q_2 \\ -q_1 & 0 & q_3 & q_1-q_3 \\ q_2 & -q_3 & 0 & -q_2+q_3 \\ q_1-q_2 & -q_1+q_3 & q_2-q_3 & 0 \end{pmatrix}$$

We finally obtain

$$I_2 = -8v/V$$

$$I_3 = -30(v/V)^2$$

$$I_4^{[3]} \cong 81.128\,(v/V)^3, \quad I_4^{[2]} \cong -120.901\,(v/V)^3, \quad I_4^{[1]} \cong 165.786\,(v/V)^3$$

(the 3 integrals $I_4^{[3]}, I_4^{[2]}, I_4^{[1]}$ are computed numerically, hence $I_4 \cong -146.92\,(v/V)^3$).

# Appendix A2

We here demonstrate that $\int \frac{\Phi(k)}{V} \frac{V d^3 \mathbf{k}}{(2\pi)^3} = -1$.

$$\int \frac{\Phi(k)}{V} \frac{V d^3 \mathbf{k}}{(2\pi)^3} = \frac{-2}{\pi} \int_0^{+\infty} \frac{1}{q} \left( \sin(qd) - qd \cos(qd) \right) dq$$

$$= \frac{-2}{\pi} \left( \int_0^{+\infty} \frac{\sin(x)}{x} dx - d \int_0^{+\infty} \cos(qd) dq \right)$$

$$= \frac{-2}{\pi} \left( \int_{-\infty}^{+\infty} \frac{\exp(ix)}{2ix} dx - d \int_{-\infty}^{+\infty} \frac{\exp(iqd)}{2} dq \right)$$

$$= \frac{-2}{\pi} \left( \frac{\pi}{2} - d\pi \delta(d) \right) = -1$$

since the diameter of the spheres is not null.